# A Data-Driven Approach for Mapping Multivariate Data to Color

Shenghui Cheng, Wei Xu, Wen Zhong and Klaus Mueller

Visual Analytics and Imaging (VAI) Lab, Computer Science Department, Stony Brook University and Brookhaven National Lab

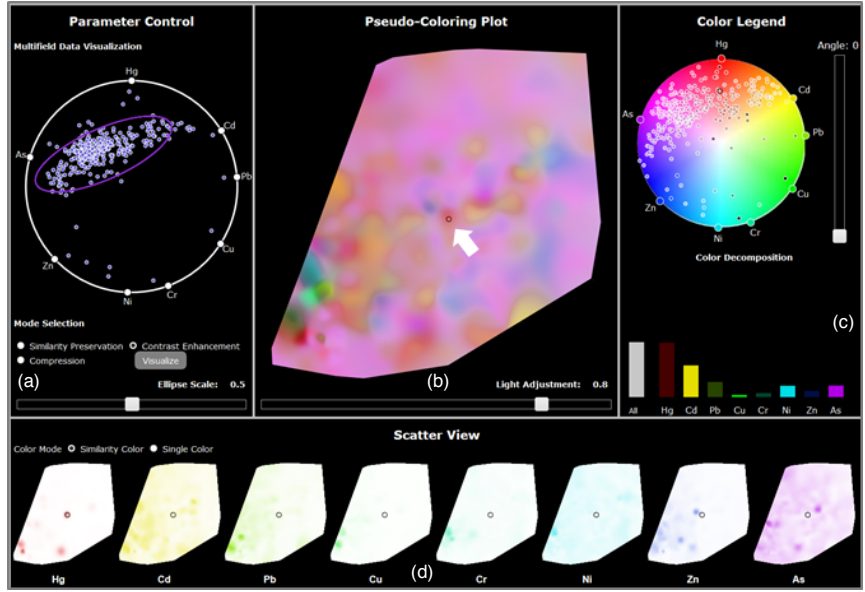

Fig. 1. The interface of our approach. The white arrow points to a small red-colored region (black dot). The color legend shows its position in the color map and the bar charts show its attributes' values. The scatter plots show the same position in different individual maps.

## ABSTRACT

A wide variety of color schemes have been devised for mapping scalar data to color. Some use the data value to index a color scale. Others assign colors to different, usually blended disjoint materials, to handle areas where materials overlap. A number of methods can map low-dimensional data to color, however, these methods do not scale to higher dimensional data. Likewise, schemes that take a more artistic approach through color mixing and the like also face limits when it comes to the number of variables they can encode. We address the challenge of mapping multivariate data to color and avoid these limitations at the same time. It is a data driven method, which first gauges the similarity of the attributes and then arranges them according to the periphery of a convex 2D color space, such as HSL. The color of a multivariate data sample is then obtained via generalized barycentric coordinate (GBC) interpolation.

## 1 INTRODUCTION

Mapping data to color has a rich history and several well-tested color schemes have emerged [1]. In this paper, we are interested in colorizing multivariate data. Here we wish to go beyond the bivariate and trivariate cases, where the two or three variables can be assigned to two or three primary colors through bilinear or barycentric interpolation respectively [3]. Our method is an automatic and data-driven method for visually encoding similarities of variables and data items.

---

\* {shecheng, wxu, wezzhong, mueller}@cs.stonybrook.edu

## 2 OUR MULTIVARIATE COLOR MAPPING SCHEME

Our color scheme is a data-driven method based on the relations of data. We first conducted the analysis to extract the similarities or distances among data items and variables. GBC plot is a typical way to visualize the relations among data items and variables, but it loses accuracy. Thus we make use of the improved GBC plot described by Cheng et al [2]. It first maps the variables at the GBC plot periphery (the vertices of the GBC – we choose a circular representation) and then map the data points into it. Since this layout scheme is an optimized approach, it is able to preserve the similarity of variables, the similarity of data points, and the similarity of data points to variables.

We illustrate our work with 300 data samples obtained at irregularly placed sensors in a city. Measured are heavy pollutant chemicals, such as "As", "Cd", "Cr", "Cu", "Hg", "Ni", "Pb", and "Zn". Fig. 1 shows a visualization of these data with our interface.

**Color Space**

After laying out the data, we seek to map the data to a proper color space. Here we wish to choose a color space that can preserve the various similarities described above. Considering the circular shape of the GBC plot, we choose the HSL color space (alternatively, an implementation with the HCL space is in progress). The HSL has a double cone topology and turns into a circle if the lightness L is fixed. This circle can be integrated into the GBC layout. Hence, the HSL color space provides a straightforward geometry for mapping. Using the GBC mapping scheme for the layout, the variables can then be positioned at the circle's boundary and consequently the samples are laid out in the circle's center. As such, the layout will result in the sample's Hue H and Saturation S. The GBC plot directly maps to the center slice of the HSL color space and given this direct mapping it is straightforward to obtain the H and S values for the geospatial map colors. For the lightness, we typically choose L=0.65 for now (Fig. 2 (a)), but we also allow users to adjust it to enhance some features.

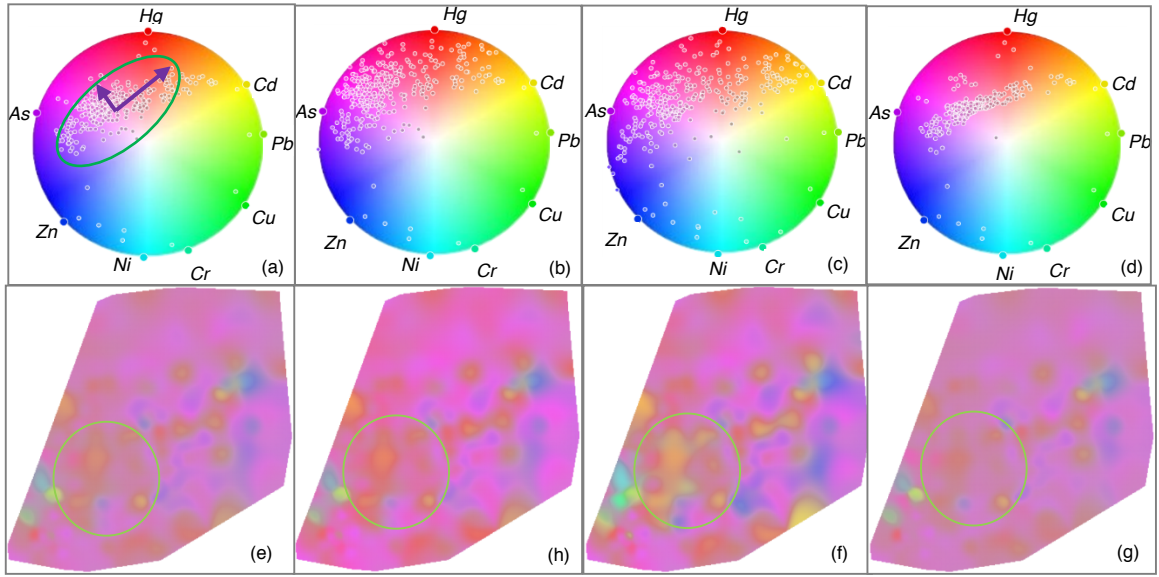

Fig 2. Different feature extraction methods and their corresponding pseudo-colored maps. (a) and (e) original coloring, (b) and (f) Color preserving enhancement coloring, (c) and (g) contrast enhancement coloring, (d) and (h) comparison compression coloring.

The GBC layout of the attributes preserves the original similarities in the data and this feature is still maintained in the HS space. By using the GBC plot and the HS space, we can map similar colors to similar attributes, and vice versa. Likewise, similar points map to similar locations in the GBC plot's interior (and the HS map) and thus will end up having similar colors. Due to the similarity-based, contextual layout of data items and variables, the GBC plot enables users to appreciate what the dominant pollutants are. It is important to realize that the HSL slice-based plot maps samples by the percentage of pollutants they have, and not by absolute value. Conversely, for drawing the colored spatial map we can map the sample weight to intensity.

After assigning colors to samples, we then generate the continuous maps based on adaptive kernel density regression (AKDE) on irregular sample points as shown in Fig. 2 (e).

**Feature Extraction**

In our example, we observe that the points mostly use colors in the upper left part of the HS space, leading to a loss of contrast. We would like to devise a method that effectively utilizes the entire HS color gamut to enhance this contrast. At the same time, there is a considerable amount of open space for outliers, which still cannot be distinguished. To explore this fused map and appreciate different features, i.e. cluster or details, from different views, we would like to give users more freedom in manipulating the color mappings. We achieve this by first identifying the elliptical distribution region of the points via principle component analysis (Fig.2 (a) purple lines with green ellipse) and then shrinking or enlarging the GBC space circle onto this ellipse warping the HS space along with it. We provide three extraction schemes: color preserving enhancement, contrast enhancement, and comparison compression coloring.

The **color preserving enhancement** method seeks to enhance contrast while keeping the original color gamut. It pushes or drags the points in the ellipse to the border of the circle, but not across the white point in the center of the HS color space. **Contrast enhancement coloring** focuses more on contrast and warps the ellipse to the full HS space. The points in the ellipse do not vary much compared to the points outside the ellipse. The **comparison compression coloring** scheme restricts the points into an even smaller region so that these points cannot take up many colors and distract the user. A comparison of these schemes is shown in Fig. 2 (b) (c) (d) and their corresponding pseudo-colored maps are shown in Fig. 2 (h) (f) and (g). Compared to Fig. 2 (e) (see green circle), we could observe more details in Fig. 2 (h) and (f), but obtain a better overview in Fig. 2 (g).

**Interface**

To better manipulate this color scheme, we integrated all into a unified interface as in Fig. 1. The configuration panel (Fig.1 (a)) allows the user to set the various system parameters, such as feature extraction mode and ellipse scale size. The main interface, *Pseudo-Coloring Plot* (Fig.1 (b)), displays the geo-map pseudo-color based on the chosen parameter settings. The *Color Legend Panel* (Fig.1 (c)) integrates the GBC plot with the color map, which doubles as a color legend. The bar chart at the bottom shows the true values of the overall pollution and attributes for the chosen point in the Pseudo-Coloring Plot. The bottom *Scatter View* displays the single-attribute heat maps one by one. The colors of these individual colored geo-maps can be set to the colors assigned for the combined heat-map, which is the color at the corresponding vertex in the GBC-based color map. Here a brighter color corresponds to lower values. Our tool is fully interactive and lends itself well for exploratory missions.

**ACKNOWLEDGEMENTS**

This research was partially supported by NSF grant IIS 1527200, the MSIP Korea, under the "IT Consilience Creative Program (ITCCP)" supervised by NIPA, and BNL LDRD grant 16-041.